# Thermodynamical approach to nanodomain tailoring in thin ferroelectric-semiconductor films.

Anna N. Morozovska[1], Eugene A. Eliseev[2].

[1]V.Lashkaryov Institute of Semiconductor Physics, National Academy of Science of Ukraine,
41, pr. Nauki, 03028 Kiev, Ukraine,
e-mail: morozo@i.com.ua

[2]Institute for Problems of Materials Science, National Academy of Science of Ukraine,
3, Krjijanovskogo, 03142 Kiev, Ukraine
e-mail: eliseev@i.com.ua

**ABSTRACT**

We propose the thermodynamical theory of nanodomain tailoring with the help of atomic force microscope electric field in thin ferroelectric-semiconductor films. We modified the existing thermodynamical models of domain formation allowing for the Debye screening, recharging of sluggish surface charge layers caused by emission current between the tip apex and sample surface.

For the first time we calculated the realistic sizes of nanodomains recorded in $BaTiO_3$, $PbZr_xTi_{1-x}O_3$ and $LiTaO_3$ ferroelectric-semiconductor thin films in contrast to the over-estimated ones obtained in the evolved approaches. We have shown that the depolarization field energy of the domain butt, Debye screening effects and field emission at high voltages lead to the essential decrease of the equilibrium domain sizes.

For the first time we obtained, that the domain radius does not decrease continuously with applied voltage decrease: the domain appears with non-zero radius at definite critical voltage applied to the tip. This result completely agrees with experimentally observed threshold domain recording in $PbZr_xTi_{1-x}O_3$ and $LiTaO_3$ thin films. Such “threshold” domain formation is similar to the first order phase transition. We hope, that our results will help one to determine the necessary recording conditions and appropriate ferroelectric medium in order to obtain the stable domains with minimum lateral size in a wide range of applied voltages.

## 1. INTRODUCTION

Ferroelectric **nanodomains** (submicron spatial regions with reversed spontaneous polarization) are caused by inhomogeneous elastic and/or electric fields with definite polarity in many ferroelectrics-semiconductors and photorefractive ferroelectrics [1], [2]. Recently one and two dimensional arrays of nanodomains have been tailored in $LiNbO_3$ [3], $LiTaO_3$ [4], $Pb(Zr,Ti)O_3$ [5], [6], [7], $BaTiO_3$ [8] and $RbTiOPO_4$ [9] ferroelectrics materials with the help of electric field caused by atomic force microscope (AFM) tip. Such nanodomain arrays could be successfully used in modern non-volatile memory devices.

So the possibilities of information recording in the ferroelectric media have been open, if only the optimization problem of high-speed writing nanodomains with high density, stability and fully controllable reversibility would be solved. First of all it is necessary to record the stable domain "dots" with minimum width in the appropriate ferroelectric medium. To realize this idea, one has to determine the dependences of domain radius on voltage applied to the AFM tip either empirically or theoretically. To our mind simple modeling seems rather urgent for the correct description of the numerous experimental results, but existing models give incomplete description of the nanodomain tailoring owing to the following reasons.

- Thermodynamical description of the nucleation processes in the perfect ferroelectric-dielectric during polarization switching proposed by Landauer [10] as far back as 1957. At present it should be applied to the domain formation with great care, because in this model the depolarization field is partially screened by the free charges on the metallic electrodes. When modeling the domain formation such upper electrode will completely screen the interior of ferroelectric from the AFM tip electric field, thus no external source would induce the polarization reversal. Only homogeneous external field can be applied to such polar dielectric covered with metallic electrodes (see Fig.1a).
- Theoretical modeling of equilibrium ferroelectric domains recorded by AFM tip proposed by Molotskii [11] considers tip electric field inside the perfect dielectric-ferroelectric with free surface, i.e. without any screening layer or upper electrode, but the semi-ellipsoidal domain depolarization field was calculated in Landauer model as if the perfect external screening expected. As a result significantly over-estimated values of domain radius was obtained at high voltages [3].
- In our recent papers [12], [13] we try to overcome the aforementioned discrepancies, taking into consideration screening layers of immovable surface charges (see Figs 1b,c) and semiconductor properties revealed by the most of ferroelectrics [2], [14], [15]. When the distances $\Delta R$ between the tip apex and sample surface is more than several nm and applied voltage is relatively low, this screening layer maintains its negative charge during the domain formation owing to the traps sluggishness. However, in the most of experiments [3]-[9] $\Delta R < 1 nm$ and thus recharging of this layer is quite possible due to the field emission of trapped carriers caused by the positively charged AFM tip. Therefore recharging of surface traps caused by the emission current should be taken into account at least at high voltages [13].

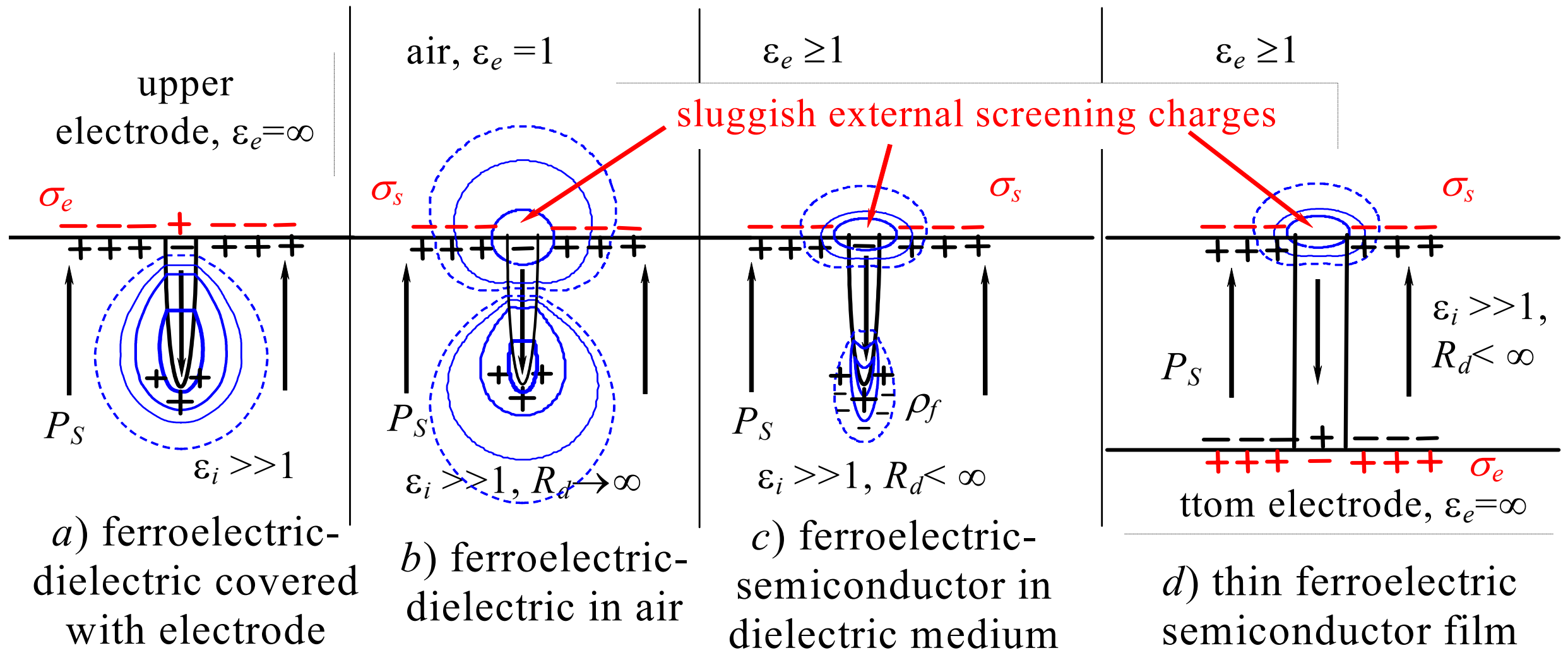

**Figure 1** Isopotential lines of depolarization field caused by nanodomain for different models: Landauer model [10] with movable charges $\sigma_e$ on the metallic electrode (a), model with surface charges $\sigma_S$ captured by traps (b), model with surface charges $\sigma_S$ and bulk free charges $\rho_f$ (c) [12], recording in thin ferroelectric-semiconductor film with bottom electrode (d).

- Nanodomain tailoring in a thin ferroelectric film with thickness no more than 100-150 nm is much more useful for applications, than domain recording in a relatively thick layer, allowing for possible miniaturization of information elements, which is extremely desirable in order to provide their figure of merits and make them preferable in comparison with existing information optical elements. Taking into account, that observed nanodomains have radiuses ~20-50 nm and cross-talk coupling between them are almost absent during their recording, the perfect possibilities of information recording by nanodomains two-dimensional arrays with extremely high density ~ (4-40)Gbits/cm$^2$ are opened [5]. As far as we know, theoretical description of nanodomain recording in thin ferroelectric films is in the initial stage: screening effects and depolarization field energy are usually either neglected or overestimated (see e.g. [16]).

In the present research for the first time we consider the influence of all aforementioned effects: depolarization field created by the domain butt, Debye screening effects, field emission at high voltages. We have shown that depolarization field of the domain butt, Debye screening and emission current lead to the essential decrease of the equilibrium domain sizes. As a result we obtained the realistic values of domain radius recorded in $Pb(Zr,Ti)O_3$ and $LiTaO_3$ thin films and demonstrate that our modeling could be really useful for the optimization of information recording.

## 2. THERMODYNAMICAL DESCRIPTION

Hereinafter we use the model of the rigid ferroelectrics with spontaneous polarization $\mathbf{P}_S$, dielectric permittivity $\varepsilon_i$, displacement $\mathbf{D} = \varepsilon_i \cdot \mathbf{E} + 4\pi\mathbf{P}_S$ and electric field $\mathbf{E} = -\nabla\varphi(\mathbf{r})$. We choose constant spontaneous polarization $+P_S$ inside and $-P_S$ outside the domain (see Fig.2).

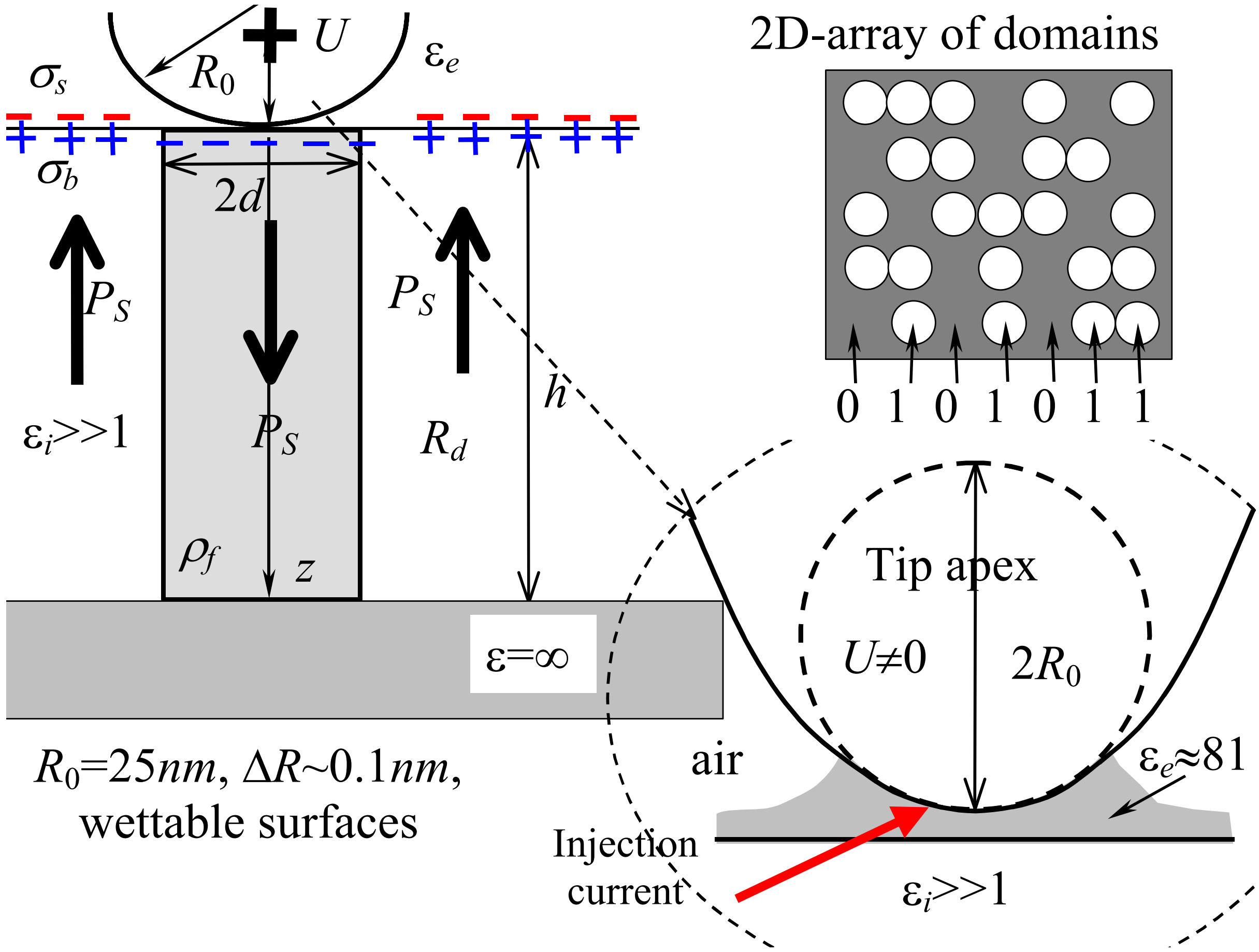

**Figure 2**. Nanodomain formation in thin ferroelectric film induced by positively charged AFM tip. Voltage $U$ is applied between the tip and ground electrode, $\Delta R$ is the distance between the tip apex and the sample surface, $R_0$ is tip radius of curvature, $z_0 = R_0 + \Delta R$ is the distance between the tip centre and the sample surface, $d$ is domain radius, $h$ is film thickness, $R_d$ is Debye screening radius, $P_S$ is spontaneous polarization, $\sigma_S$ is surface charges captured on the trap levels, $\sigma_b$ is bound charges related to $P_S$ discontinuity.

In ferroelectric-semiconductor the Schottky barriers, band bending, field effects as well as Debye screening cause surface charge layer that effectively shields the interior of the sample from the strong homogeneous depolarization field $E_d = -4\pi P_S$ [2]. Surface charges $\sigma_S = -P_S$ are captured on the sluggish trap levels before the domain formation [17] (see red minuses in the Fig.2).

For description of the charged tip electric field we use the spherical model [18], in which the tip is represented by the charged sphere with radius $R_0$ located at the distance $\Delta R$ from the sample surface. The voltage $U$ is applied between the tip and bottom electrode. The validity of such assumptions for dielectric sample and more sophisticated models are discussed in [18]. Usually $\Delta R < 1nm$ [3]-[9], sample thickness $h \sim (50nm - 500nm)$, thus the field emission is quite possible at voltages $U \geq U_m$. Keeping in mind approach proposed in [19] for current-voltage characteristics of ferroelectric tunnel junctions and interface screening model evolved in [20], we assume that emission current $J_e \sim \exp(-U_m/U)$. At applied voltages $U \geq U_m$ the amount of emitted carriers are quite enough to completely screen the reversed polarization of the domain butt-end [21], i.e. $\sigma_S \to +P_S$. So, the equilibrium surface charge density $\sigma_S$ has the form:

$$\sigma_S = \begin{cases} -P_S + 2P_S \exp(-U_m/U) & \sqrt{x^2+y^2} < d \\ -P_S & \sqrt{x^2+y^2} > d \end{cases} \tag{1}$$

The activation voltage $U_m$ complexly depends over the distance $\Delta R$, sample thickness $h$, Debye screening radius $R_d$, dielectric permittivity $\varepsilon_i$ and other sample-tip material parameters. It seems reasonable that $U_m \sim E_m h$, where $E_m$ could be considered as the threshold field.

Another important experimental fact should be taken into consideration for correct theoretical description of nanodomains tailoring using AFM tip [6], [16], [22]. Namely, the Durkan et al. [6] reported about the layer of the adsorbed water located below the tip apex. Molotskii [16] assumes that a water meniscus appears between the AFM tip apex and a sample surface due to the air humidity. For instance, scanning tunnelling microscopy measurements on Ti show that within the interval 20%–50% humidity water layer thickness increases linearly from 50 to 100 nm (see [22] and ref. therein). The authors argue that this strong water adsorption is based on an oxide layer on the Ti surface. Thus, it seems quite probable that on the $BaTiO_3$, $PbTiO_3$ or other polar oxide surface thin water layers are quite possible. Moreover, if AFM tip is wettable, its apex with curvature 25-50 nm can be completely covered by water [23]. To our mind, all these effects can be easily explained taking into account that dipolar water molecules are pulled into inhomogeneous electric field [24] and then condensates in the area of tip apex - domain surface (see Fig. 2). Hereinafter we regard, that the region between the tip apex and domain surface has effective dielectric permittivity $\varepsilon_e$.

Inside an extrinsic semiconductor $\left|Ne\varphi(\mathbf{r})/k_B T\right| << 1$ and thus the screening of electric field $\mathbf{E} = -\nabla\varphi(\mathbf{r})$ is realized by free charges with bulk density $\rho_f(\mathbf{r}) \approx -\varepsilon_i\,\varphi(\mathbf{r})/4\pi R_d^2$ and Debye screening radius $R_d$ (see Appendix A). The spatial distribution of the electrostatic potential should be determined from the Maxwell equation $\varepsilon_i\,\Delta\varphi(\mathbf{r}) = -4\pi\rho_f(\mathbf{r})$ supplemented by the interfacial conditions $D_{n\,\mathrm{int}} = D_{n\,ext}$ on the domain surface Σ, $D_{n\,ext} - D_{n\,\mathrm{int}} = 4\pi\sigma_S$ on the free surface $z = 0$ and potential disappearance at the bottom electrode surface $z = h$. Thus we obtain the following boundary problem:

$$\begin{gathered}
\Delta\varphi_0(\mathbf{r}) = 0, \qquad z \le 0, \\
\left.\varphi_0\right|_{\mathbf{r}\in AFM_{tip}} = U, \qquad \varphi_0(z=0) = \varphi(z=0), \\
\left.\left(\varepsilon_e\frac{\partial\varphi_0}{\partial z} - \varepsilon_i\frac{\partial\varphi}{\partial z}\right)\right|_{z=0} = \begin{cases} 4\pi(\sigma_S - P_S), & \sqrt{x^2+y^2} < d \\ 0, & \sqrt{x^2+y^2} > d \end{cases} \\
\Delta\varphi(\mathbf{r}) - \frac{\varphi(\mathbf{r})}{R_d^2} = 0, \qquad z \ge 0, \\
\left.\varepsilon_i\left(\frac{\partial\varphi_{\mathrm{int}}}{\partial n} - \frac{\partial\varphi_{ext}}{\partial n}\right)\right|_{\Sigma} = \left.8\pi(\mathbf{P}_S\,\mathbf{n})\right|_{\Sigma}, \quad \varphi(z=h) = 0
\end{gathered} \tag{2}$$

In order to apply all the following results to the anisotropic semiconductor one can make the substitution: $z \to z\sqrt{\varepsilon_a/\varepsilon_c}$, $\varepsilon_i \to \sqrt{\varepsilon_c\varepsilon_a}$, $R_d^2 = \dfrac{\varepsilon_a k_B T}{4\pi e^2 n_d}$. Here $\varepsilon_a$ and $\varepsilon_c$ are anisotropic dielectric permittivity values perpendicular and along the polar axis *z*.

These domains usually grow through the thin film (see e.g. [7] and Fig.2). Our preliminary calculations proved, that equilibrium domain length $l$ nearly always is much greater than film thickness $h \sim 100\,nm$, i.e. they usually grow through the film and rich the bottom electrode. For example domain length $l \ge 5\,\mu m$ for a thick $LiNbO_3$ sample corresponds to the voltage $U \ge 35V$. So, nanodomains recorded in thin films have rather cylindrical shape than the semi-ellipsoidal one. Thus hereinafter we put $l = h$.

The electrostatic potential $\varphi(\mathbf{r})$ of cylindrical domain grown through the thin film and its thermodynamic potential $\Phi(d,U)$ are derived in Appendix A under the typical recording condition $\Delta R << R_0$. Our calculations have shown, that interaction energy $\Phi_U(d)$ between the charged tip apex and the domain decreases with film thickness decrease:

$$\Phi_U(d) = \frac{\sqrt{\varepsilon_a\varepsilon_c}+\varepsilon_e}{\sqrt{\varepsilon_a\varepsilon_c}-\varepsilon_e}\ln\left(\frac{\sqrt{\varepsilon_a\varepsilon_c}+\varepsilon_e}{2\varepsilon_e}\right)\cdot\frac{4\pi\varepsilon_e(\sigma_S - P_S)\cdot UR_0\cdot hR_d\left(\sqrt{z_0^2+d^2}-z_0\right)}{\left(\sqrt{\varepsilon_a\varepsilon_c}+\varepsilon_e\right)hR_d + \sqrt{\varepsilon_a\varepsilon_c}\,(R_d+2h)\sqrt{z_0^2+d^2}} \tag{3}$$

The depolarization field energy $\Phi_D(d)$ caused by polarization reversal inside the cylindrical domain grown through the film also decreases with film thickness decreases with film thickness decrease:

$$\Phi_D(d) \approx \begin{cases} \dfrac{\pi^2(\sigma_S - P_S)^2 d^3 h R_d}{h\left(\sqrt{\varepsilon_a\varepsilon_c}\, d + \left(\sqrt{\varepsilon_a\varepsilon_c} + \varepsilon_e\right)3\pi R_d/16\right) + \sqrt{\varepsilon_a\varepsilon_c}\, R_d d/2} & at \quad d \gtrsim R_0/2 \\ \pi^2(\sigma_S - P_S)^2 d^2 \dfrac{R_d\left(2\varepsilon_e R_d + \sqrt{\varepsilon_a\varepsilon_c}\Delta R\right)}{\left(\varepsilon_e R_d + \sqrt{\varepsilon_a\varepsilon_c}\,\Delta R\right)^2}\Delta R & at \quad d << R_0, \quad \Delta R << h \end{cases} \tag{4}$$

It should be noted that the similar problem (cylindrical microdomain growing through the thin film) was considered by Molotskii in [16]. However the depolarization field energy dependence on the film thickness and domain radius was chosen as $d^2 h$ in the case $h >> d$. As it follows from our energy (4) the dependence $d^2 h$ can be expected in the opposite case $h << d$ (see (A.12)). In the case $h >> d$, $R_d >> d$ our calculations give $\Phi_D(d) \sim d^3$. Since the coefficients of proportionality in (4) and in the expressions (10), (13) used in [16] are same order of magnitude, the depolarization field energy [16] for the films with thickness $h >> d$ is strongly overestimated. Moreover for the most of the experiments $d \sim h$ and approximation like (4) is more useful than limiting expressions like (A.12).

The surface energy of domain wall has the form:

$$\Phi_C(d) = 2\pi\psi_S d\, h \tag{5}$$

We regard domain walls as infinitely thin, with homogeneous surface energy density $\psi_S$. This assumption used by many authors [10], [11], [16], [25] was confirmed both by the modern first principle calculations (e.g. [26], [27], [28]) and recent experimental data (e.g. [29], [30]), which revealed that the thickness of the domain walls in perovskite ferroelectrics is about several lattice constants. The value $\psi_S$ can be calculated from the first principles [26], [27], [28] or measured experimentally from the rate of domain wall motion, their thickness or curvature (e.g. [31], [32]). Depending on the sample preparation and experiment conditions the measured energy $\psi_S$ differs from its calculated equilibrium value in several to several hundreds of times (e.g. compare extracted from domain walls motion values 9-35 mJ/m$^2$ for $LiTaO_3$ [31] with the reconstructed ones from the domain wall curvature 200-400 mJ/m$^2$ [32]). High applied voltages as well as nonequilibrium carriers (emitted, photo-ionized, UV- or thermo-activated ones) could significantly increase the domain wall velocity [2], whereas the equilibrium conductivity, that determines Debye screening radius $R_d$, virtually does not effect on the domain walls structure [2]. Domain wall kinetic energy is consumed to overcome the wall pinning under the domain growth, thus unavoidable growth defects pin the domain wall and significantly increase domain wall relaxation time. Therefore $\psi_S$ values for perfect samples are essentially smaller than for the imperfect ones [33]. Note, that the authors of [28] showed, that the free energy $\psi_S$ is usually smaller than the domain wall energy, because it includes entropic contribution. Taking into account aforementioned facts, we consider effective surface energy $\psi_S$ as a fitting parameter varied in the range 5-500 mJ/m$^2$.

Taking into account (1)-(5), free energy of the nanodomain recorded in thin film acquires the form:

$$\Phi(d,U) \approx \Phi_U(d) + \Phi_D(d) + 2\pi\psi_S d\, h \tag{6}$$

Similarly to the case of semi-ellipsoidal domains recorded in the thick films [13], it is appeared that free energy (6) has the absolute minimum $\Phi_{\min}(d,U) < 0$ at $d = d_{\min}$ only when $U \geq U_{cr}$ (see Fig. 3).

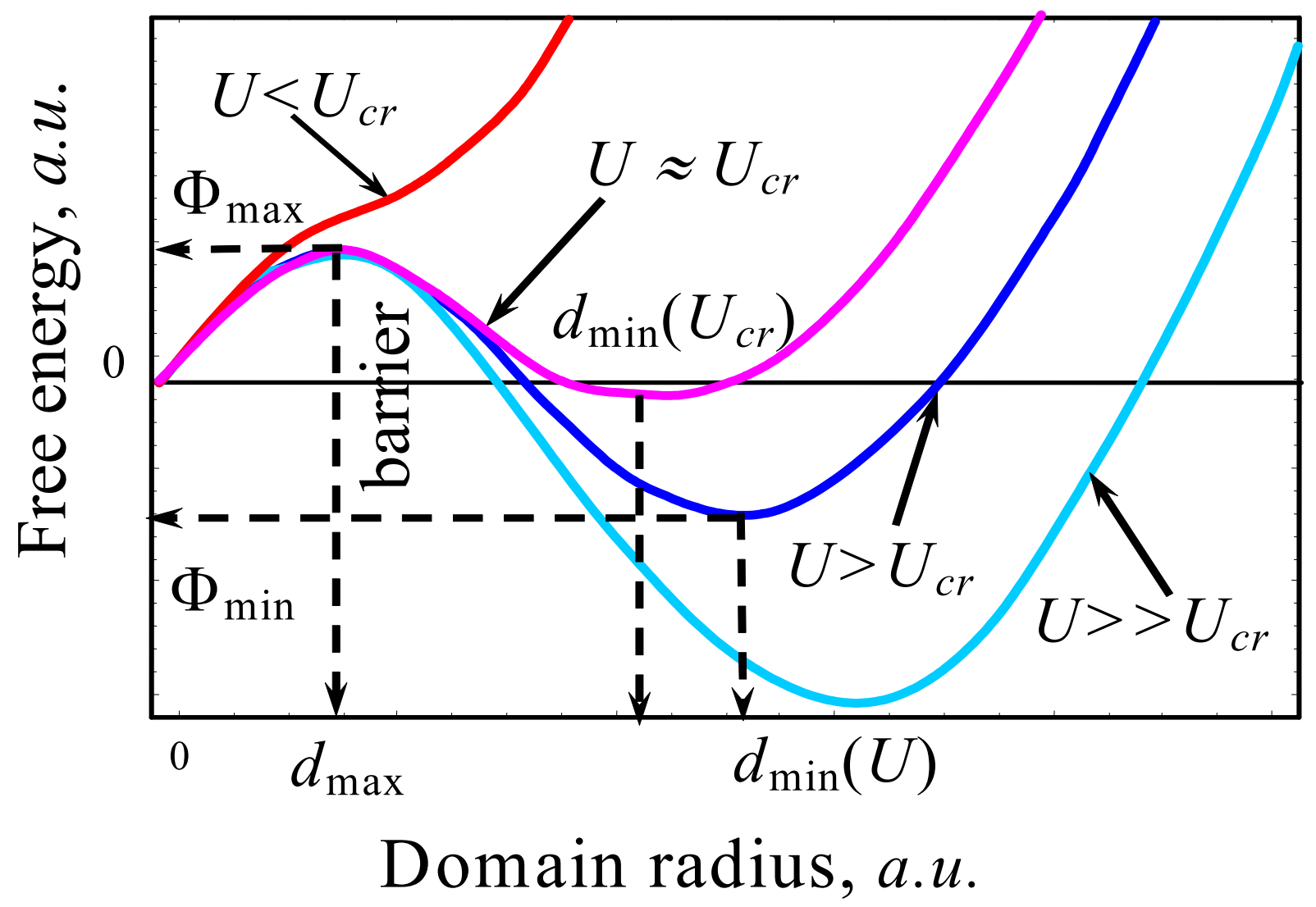

**Figure 3.** Free energy dependence on the domain radius for the different applied voltage values *U*.

At lower voltages $U < U_{cr}$ the domain formation becomes energetically impossible. From the condition $\Phi_{\min}(d_{\min}, U) = 0$ we derived the following expressions for $U_{cr}$ and $d_{\min}(U_{cr})$ determination:

$$U_{cr}(h) \approx \sqrt{\frac{32\psi_S}{3}\frac{\sqrt{\varepsilon_a\varepsilon_c}R_0 + \left(\sqrt{\varepsilon_a\varepsilon_c} + \varepsilon_e\right)h}{\sqrt{\left(\sqrt{\varepsilon_a\varepsilon_c} + \varepsilon_e\right)h}}\left(\varepsilon_e\frac{\sqrt{\varepsilon_a\varepsilon_c} + \varepsilon_e}{\sqrt{\varepsilon_a\varepsilon_c} - \varepsilon_e}\ln\left(\frac{\sqrt{\varepsilon_a\varepsilon_c} + \varepsilon_e}{2\varepsilon_e}\right)\right)^{-1}} \tag{7}$$

$$d_{\min}(U_{cr}) \approx \sqrt{\frac{3\psi_S\left(\sqrt{\varepsilon_a\varepsilon_c} + \varepsilon_e\right)}{8\left(\sigma_S(U_{cr}) - P_S\right)^2}\cdot h} \tag{8}$$

Expressions (7)-(8) were derived for the case $d_{\min}(U_{cr}) < R_0$, $d_{\min}(U_{cr}) << h$, $d_{\min}(U_{cr}) << R_d$ .

The value $U_{cr}$ determines the point where the homogeneous polarization distribution becomes absolutely unstable. Such "threshold" domain formation is similar to the well-known first order phase transition. This result seems quite reasonable, because the most of stable ferroelectric nanodomains were recorded only above some critical voltage $U_{cr} \sim (3-6)V$ (see e.g. [4], [5], [6]). It should be noted that the minimal value of domain radius (8) has the same dependence on the film thickness and domain wall energy as the equilibrium lateral size of the periodic stripe domain structure (see [1] and references therein).

In the majority of experiments [3], [4], reversed nanodomains remain unchangeable during many days and weeks. This fact proves their stability and is extremely useful for applications. It has the following explanation within the framework of the proposed model. When applied voltage $U$ is turned off, the external field disappears as proportional to $U$ and the domain depolarization field $\mathbf{E}_D = -\nabla\varphi_D \sim (\sigma_S - P_S)$ (see (A.7))) vanishes due to the final recharging of the surface traps with $\sigma_S \to P_S$. The remained positive surface energy $\Phi_C(d) \sim \psi_S d\, h$ tends to reverse the domain. In order to move, the domain wall has to overcome the pinning caused by defects, but the calculated tension field $E_W \cong \psi_S / P_S\, d \sim 100\, kV/cm$ appears too small in comparison with thermodynamic coercive field $E_C \cong 4\pi P_S / \sqrt{\varepsilon_a \varepsilon_c} \sim (250-2700) kV/cm$ regarded necessary for the domain re-polarization. Thus the domain walls pinning does provide the recorded domain stability, surely it also determines the domain growth kinetics [5].

Let us discuss the peculiarities of possible inter-influence between domains during their recording. In accordance with conventional rules of binary information recording the distance between the centers of neighbour domain dots must be about $2d$ (see right inset in the Fig. 2). Therefore neighbour nanodomains can interact via their depolarization fields $E_D$. If the time interval $t$ between recording neighbour domains is enough for complete recharging of the surface traps, depolarization field vanishes and domain inter-influence is absent. Thus, during such "**equilibrium slow**" recording it is possible to obtain independent nanodomains with minimum radius $d_{\min}$ at $U \approx U_{cr}$ (see (8)). However, the faster continuous recording of independent nanodomains seems rather useful for applications. Taking into account that the system relaxation time $\Gamma$ quickly decreases with $U$ increase above $U_{cr}$, the fastest recording can be realized in the case $U >> U_{cr}$ and $U >> U_m$, i.e. when $\exp(-U_m/U) \to 1$, $\sigma_S \to P_S$ and so $E_D \to 0$. Taking into account that domain radius $d$ increases with voltage $U$ increase, usually $d >> d_{\min}$ at $U >> U_{cr}$. Thus in order to record 2D-arrays of almost independent nanodomains with $d \gtrsim d_{\min}$ in the continuous mode, the optimization of recording conditions is necessary.

Such optimization can be realized when nanodomain thermodynamic potential $\Phi(d,U)$ is known (e.g. it can be found from "**equilibrium**" experimental data). For instance, at chosen recording time $t$ and fixed film thickness $h$ one can calculate the optimal voltage $U_{opt}$ from the conventional relaxation equation [16]. In the next section we'll show, that our thermodynamic potential (6) can be unambiguously reconstructed from experimental dependence $d_{\min}(U)$.

## 3. NANODOMAIN RECORDING IN THIN FERROELECTRIC FILMS

### 3.1. Modeling of equilibrium domain sizes in typical ferroelectric-semiconductors.

The dependences of equilibrium domain radius $d$ over applied voltage $U$ for different values of Debye screening radius $R_d$ and film thickness $h$ are shown in the Figs. 4. The chosen parameters are typical for $BaTiO_3$ crystals.

The depolarization field energy (4) of the domain butt essentially decreases the domain sizes even at low voltages [12]. We would like to underline, that Debye screening not only decreases depolarization field inside the domain, but also it shields the interior of the sample from the AFM tip electric field. As a result, Debye screening radius decrease leads to the essential decrease of the equilibrium domain sizes (see the lowest curves in the Fig. 4a). At small radius of curvature $R_0 << R_d$ and distance $\Delta R << R_0$ the critical voltage $U_{cr}$ are almost independent over $R_d$ value, e.g. $U_{cr} \approx 2.5\,V$ corresponds to $R_d \geq 100\,nm$ and $h = 100\,nm$ (see Fig. 4a and (7)). The domain radius $d$ essentially decreases and critical voltage $U_{cr}$ essentially increases under $h$ increase (see Fig. 4b).

In general case carriers emission leads to the essential decrease of the of the domain sizes at high voltages, namely the domain growth stops at $U >> U_m$ (see saturated curves for $U > 30\,V$). Really at $U_m \to 0$ surface density $\sigma_S \to P_S$ and so emitted carriers fully compensate not only the depolarization field caused by the reversed polarization of the domain butt, but they simultaneously screen the charged tip electric field, which is the reason of the domain formation.

Note, that nanodomains with radius about 30nm were recorded at applied voltages about 80V in $BaTiO_3$ films [8]. Unfortunately the authors did not report about the tip curvature $R_0$, so we could not present the quantitative comparison with [8].

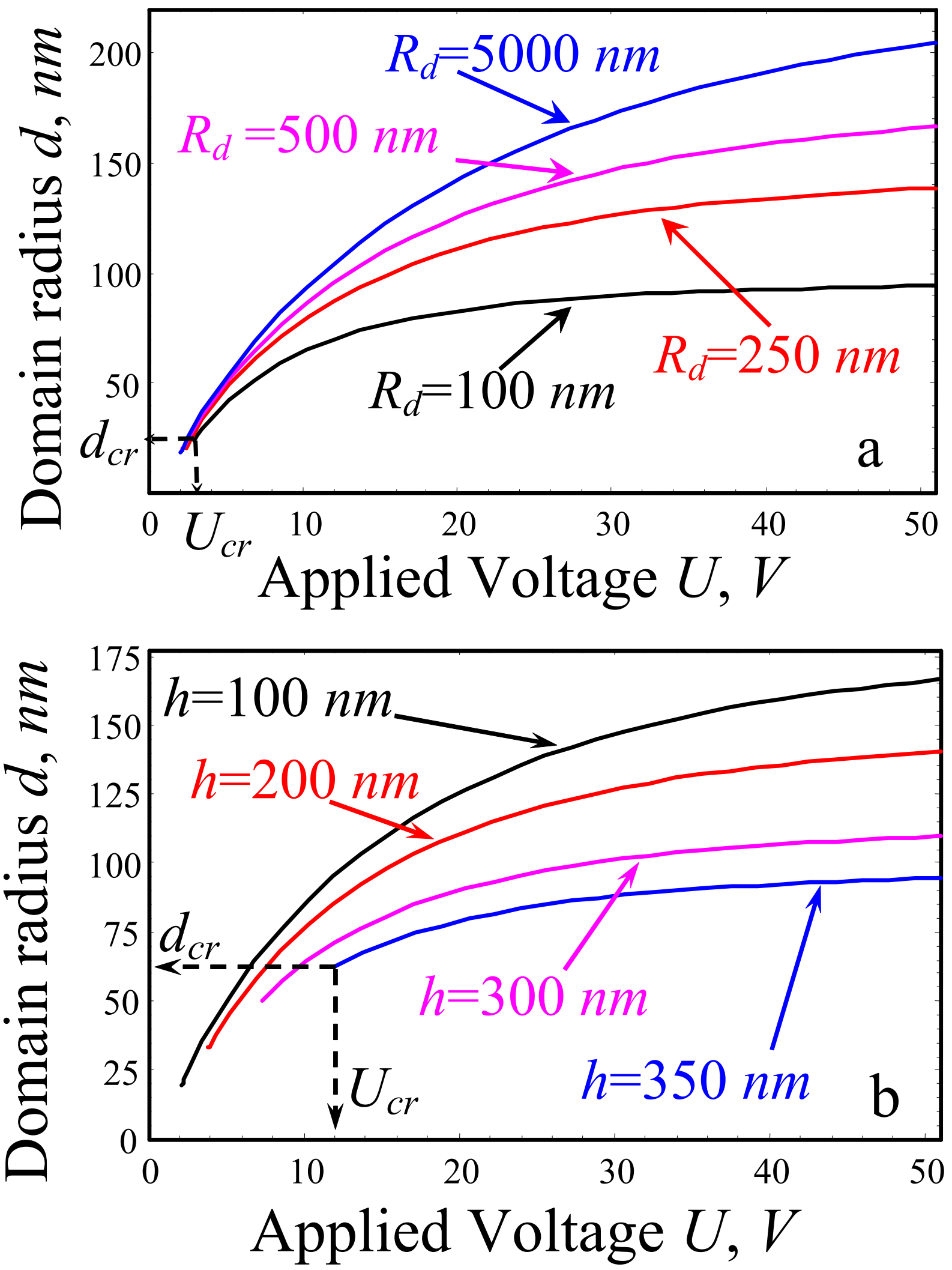

**Figure 4.** The dependences of equilibrium domain radius $d$ over applied voltage $U$ for different Debye screening radius $R_d$ at $h = 100\,nm$ (a) and film thickness $h$ and $R_d = 500\,nm$ (b). We used typical for $BaTiO_3$ parameters [1] $P_S \approx 26\,\mu C/cm^2$, $\varepsilon_a = 2000$, $\varepsilon_c = 120$, $\psi_S \approx 8\,mJ/m^2$ [28] and $\varepsilon_e = 81$, $R_0 = 25\,nm$, $\Delta R \le 1\ nm$, $U_m = 1\ V$.

### 3.2. Nanodomain recording in thin Pb(Zr,Ti)O$_3$ films

Let us apply our theoretical results (6)-(8) to the nanodomain formation in $PbZr_{0.2}Ti_{0.8}O_3$ thin film on $SrTiO_3$ substrate. In experiment [5] AFM tip radius $R_0 = (20-50)\,nm$; the distance $\Delta R \le 0.3\,nm$ was difficult to control due to the surface roughness 0.2-0.3 nm; applied voltage pulse values varied in the range $U = (4-12)\,V$; pulse duration was about $0.1\,s$. The comparison of experimental results [5] with our calculations is presented in the Fig. 5.

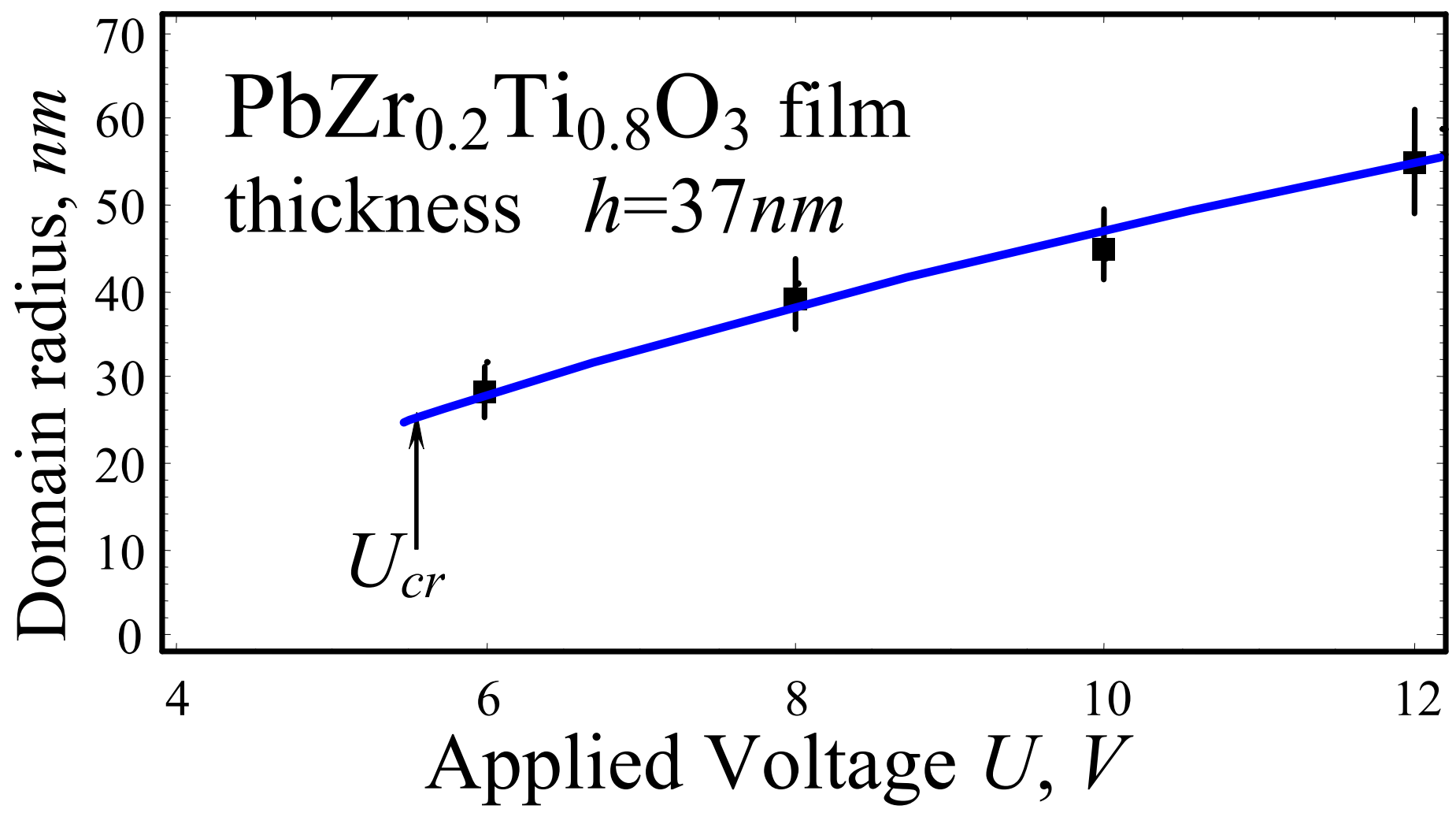

**Figure 5.** The dependence of nanodomain radius *d* over applied voltage *U*. Squares with error bars are experimental data from [5] for nanodomains recorded in thin $PbZr_{0.2}Ti_{0.8}O_3$ film with thickness $h = 37\,nm$; $R_0 \approx (20-50)nm$, $\Delta R \le 0.3\,nm$ was difficult to control due to the surface roughness 0.2-0.3 nm. Best fitting was obtained at $\varepsilon_a = 143$, $\varepsilon_c = 86$, $P_S \approx 70\,\mu C/cm^2$, $\psi_S \approx 140\,mJ/m^2$, $\varepsilon_e = 81$, $R_d = 1\,\mu m$, $R_0 \approx 35\,nm$, activation voltage $U_m \approx 1.5\,V$.

The spontaneous polarization $P_S$ and dielectric permittivity components $\varepsilon_{a,c}$ were calculated with the help of coefficients [34] for the bulk $PbZr_{0.2}Ti_{0.8}O_3$. It is appeared that the value $P_S \approx 70\,\mu C/cm^2$ coincides with the one extracted from the *a/c* ratio measured in the same thin films [29]. The fitting value of surface free energy $\psi_S \approx 140\,mJ/m^2$ is the same order as the values of $PbTiO_3$ domain wall energy $(100-200)mJ/m^2$ calculated by [26], [27] from the first-principles. Thus, our fitting makes it possible to reconstruct the surface free energy value in $PbZr_{0.2}Ti_{0.8}O_3$ thin film, which is in a good agreement with first-principle calculations [26], [27].

Note, that the fitting value of the threshold field $E_m \approx U_m/h = 405\,kV/cm$ is in a qualitative agreement with the quasi-static coercive field values $E_C \sim 100\,kV/cm$ typical for thin PZT films (see e.g. [14] and ref therein). The value $E_m$ is much higher than air breakdown field $E_b \approx 30\,kV/cm$, as it should be expected.

The dependence of the free energy (6) on the domain radius for different values of the applied voltage *U* is represented in the Fig 6 for $PbZr_{0.2}Ti_{0.8}O_3$ parameters. It is clear from the figure, that free energy has no minimum at small voltages $U \le 4V$. With the voltage increase $U \ge 6V$ the absolute minimum $\Phi_{\min}(d) < 0$ appears at definite value of domain radius $d \approx 20\,nm$. With the further increase of applied voltage the equilibrium domain radius rapidly increases and the stable domain appears (see curves for $U \geq 8\ V$).

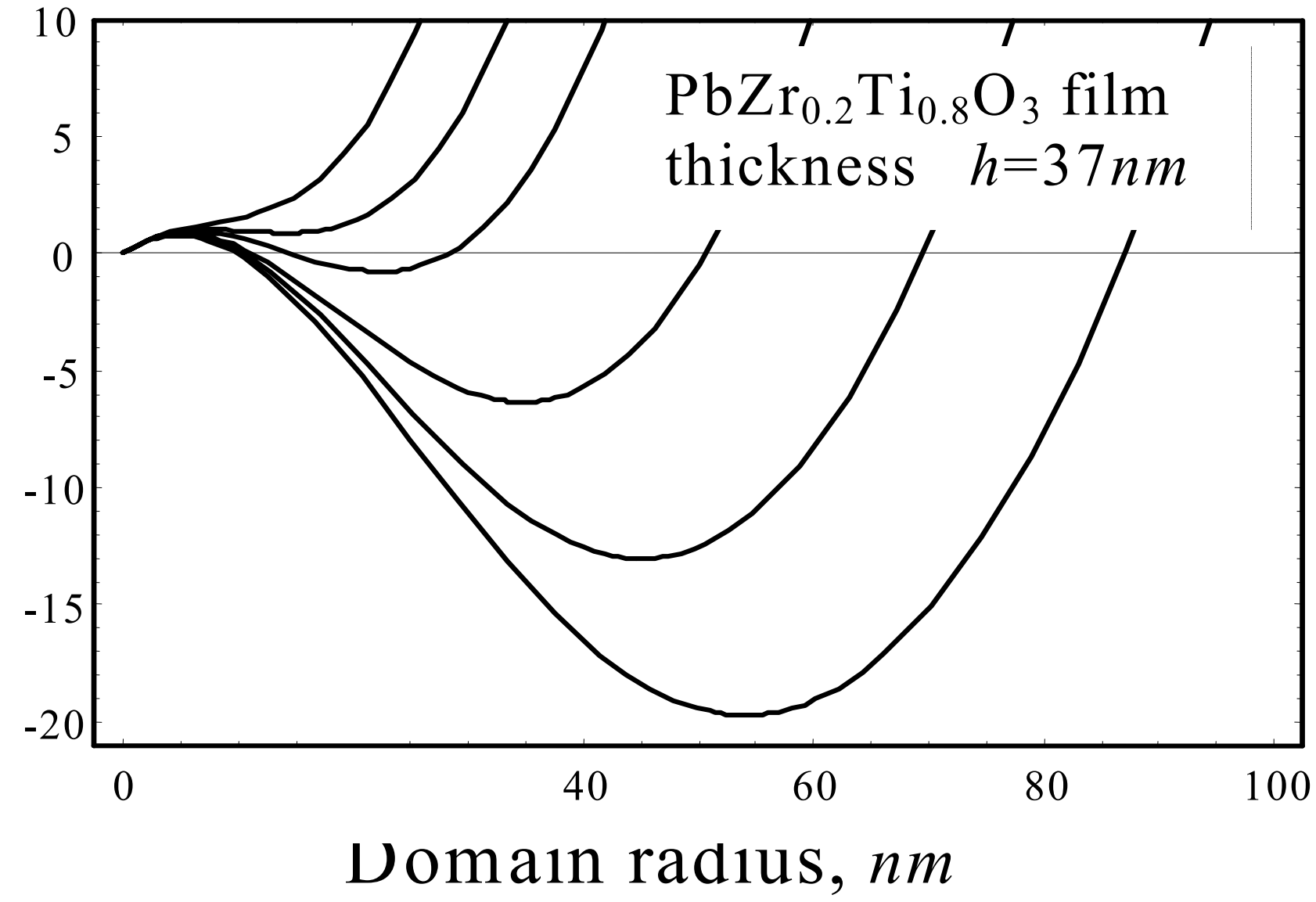

**Figure 6.** Free energy dependence on the domain radius for the different $U$ values and the same other parameters as in the Fig. 5.

It is clear from the Fig. 6 that there is a potential barrier between $d = 0$ (domain is absent) and $d \neq 0$ (domain exists). The relative height of this barrier sharply increases with applied voltage decrease (compare curves for $U = 6\ V$ and $U = 12\ V$). That is why the domain radius does not decrease continuously with voltage decrease: the domain appears with non-zero radius $d_{cr}$ at $U > U_{cr}$ (see (7)-(8)). It is seen from the figure that the critical voltage of nanodomain appearance is about 5 $V$, which is in a good agreement with the value reported by the authors [5].

The authors [6] recorded nanodomains with radius $d = (50 - 200)nm$ in lead zirconate-titanate sol-gel film with thickness $h = 190\,nm$. Applied voltage pulse values varied in the range $U = (3 - 8)\,V$ with duration about $0.1\,s$. The comparison of their experimental results and our calculations is presented in the Fig.7.

The small free energy, spontaneous polarization and high permittivity fitting values are the same order as the $BaTiO_3$ parameters (compare with Fig. 4). Such parameters are typical for soften $Pb(Zr,Ti)O_3$ nano-grained films with chemical composition near the morphotropic boundary. Really, the authors [6] reported about grain sizes $\leq 10\,nm$. How could conventional Landauer approach, evolved for the perfect ferroelectric material, be applied to the nano-grained sample? In order to describe this experiment, the author [16] used Landauer free energy (without domain butt depolarization energy) and fitting value $P_S \leq 10\,\mu C/cm^2$ "averaged over the grain orientations". To our mind, such solution needs additional basing. In our recent papers [14], [15], we modified Landau-Khalatnikov approach for inhomogeneous ferroelectrics and shown, that polar and dielectric properties of such materials can be described by three coupled equations. Within the framework of our model [14], [15] all the parameters $\langle P_S \rangle$, $\langle \varepsilon_{a,c} \rangle$ etc are renormalized by the bulk inhomogeneities, e.g. the averaged displacement obeys the modified Landau-Khalatnikov equation with renormalized coefficients. Allowing for these results, we use free energy (6) with renormalized values $P_S$ and $\psi_S$. Note, that our fitting value of the threshold field $E_m \approx U_m/h = 53\,kV/cm$ is in a reasonable agreement with the quasi-static coercive field values $E_C \sim (20-60)\,kV/cm$ typical for soften Pb(Zr,Ti)$O_3$ ceramics (see e.g. [14]).

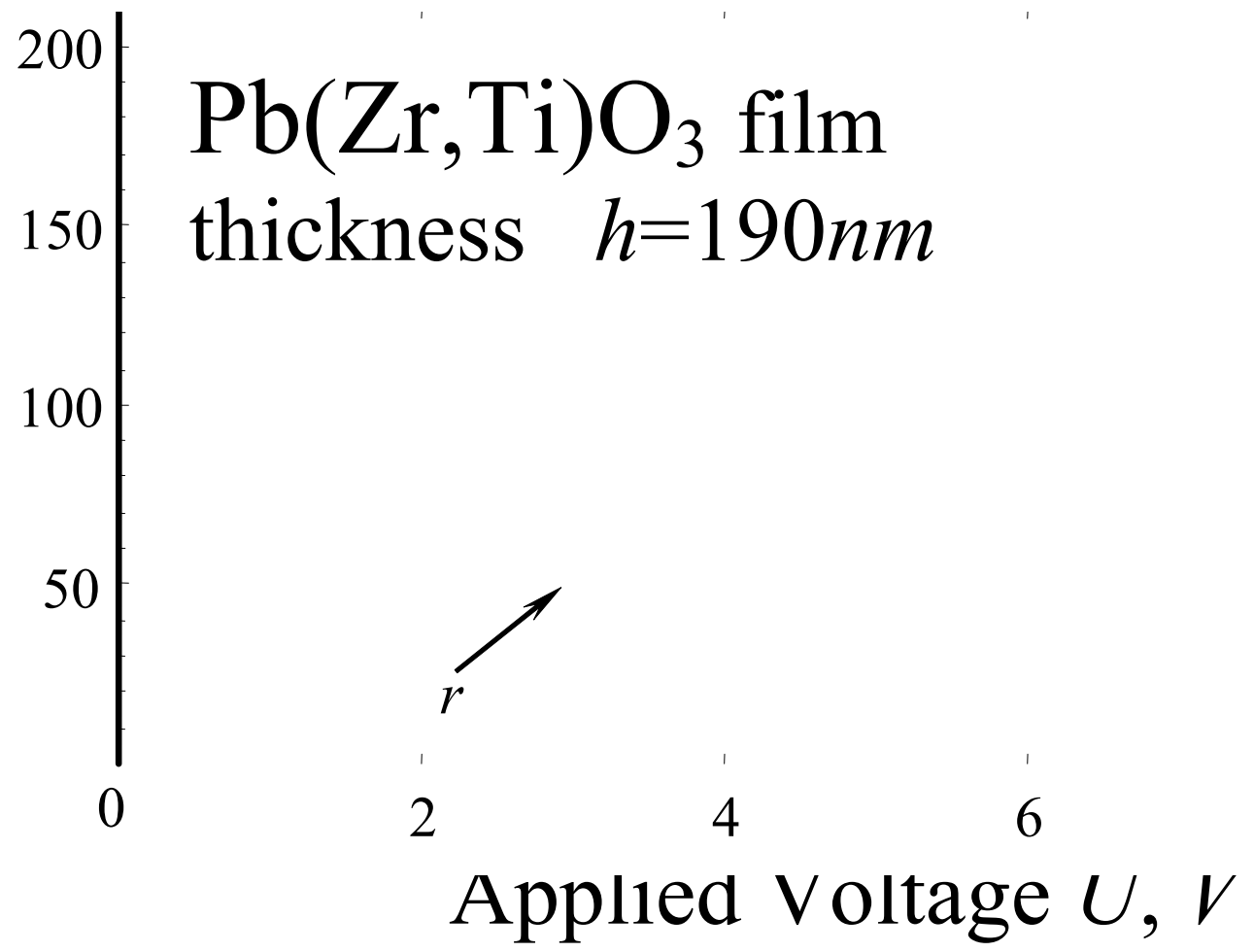

**Figure 7.** The dependence of nanodomain radius *d* over applied voltage *U*. Squares are experimental data from [6] for Pb(Zr,Ti)$O_3$ sol-gel film with thickness $h = 190\,nm$ and $\varepsilon_a = 1500$, $\varepsilon_c = 1500$, $\varepsilon_e = 81$, $P_S \approx (34-54)\mu C/cm^2$, $R_0 \approx 100\,nm$, $\Delta R \approx 6\,nm$. Best fitting was obtained at $R_d = 1\,\mu m$, $\psi_S \approx 4\,mJ/m^2$, $P_S \approx 33\,\mu C/cm^2$, $U_m \approx 1\,V$, $U_{cr}$ is about $3V$.

It is clear from the Figs. 5 and 7, that we obtained rather well quantitative agreement between our modeling and domain radiuses recorded in thin Pb(Zr,Ti)$O_3$ films at different applied voltages.

### 3.3. Nanodomain recording in thin LiTaO$_3$ films

Let us apply our theoretical results (4)-(6) to the nanodomain formation in congruent LiTa$O_3$ thin films using high-voltage AFM. In experiment [4] AFM tip radius was $R_0 = 25nm$, maximum applied voltage pulse value $U_{\max} = 11\,V$ with duration up to $10^{-4}\,s$, and LiTa$O_3$ thin films have thickness $h = 55 \div 83\,nm$. The free energy density $\psi_S \approx (60-400) mJ/m^2$ was calculated in [33] for $LiTaO_3$ at $T$=300K, $\varepsilon_a = 51$, $\varepsilon_c = 45$, $P_S \approx 55\mu C/cm^2$. The comparison of experimental results and our calculations is represented in the Figs. 8-10.

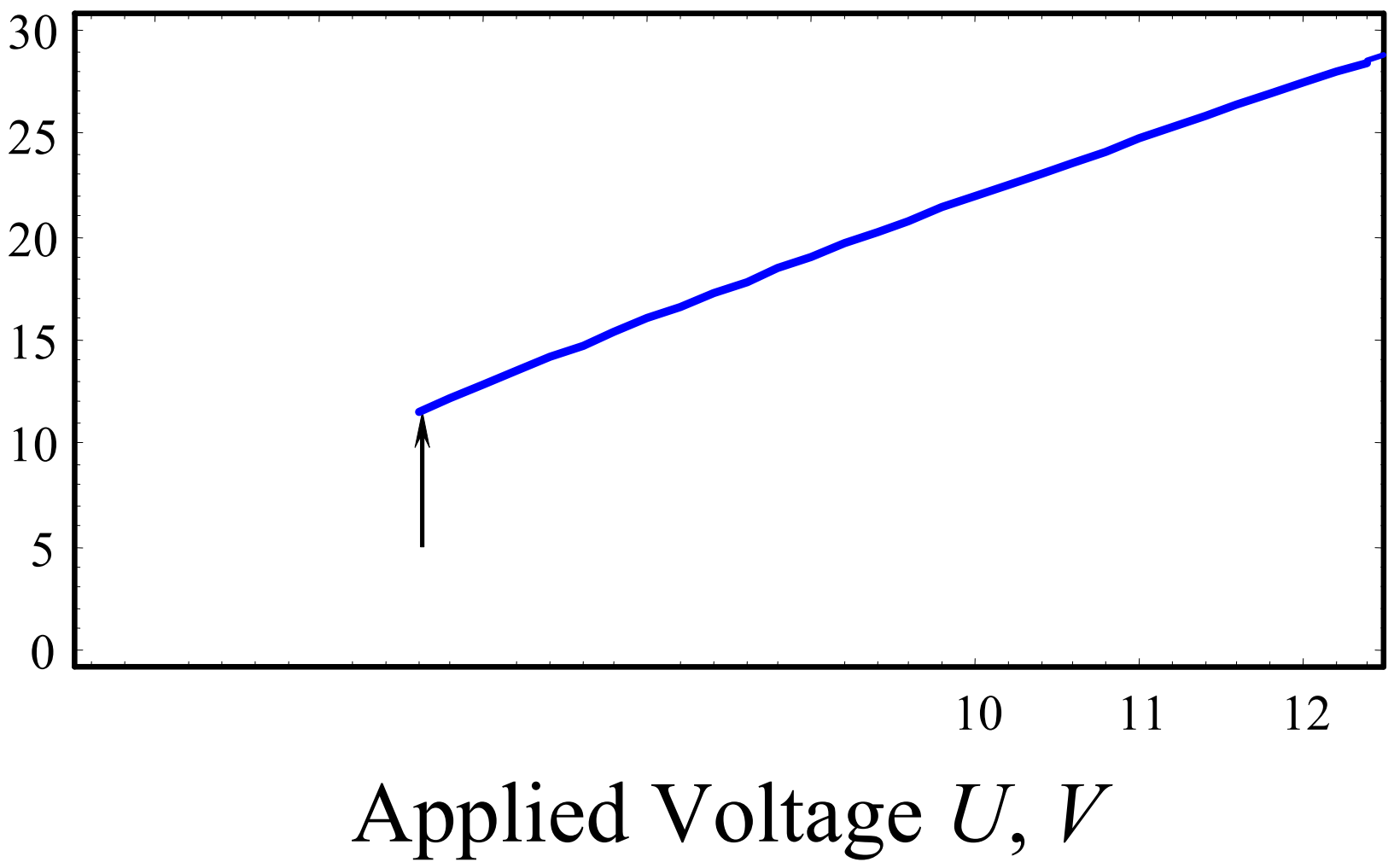

**Figure 8.** The dependence of nanodomain radius *d* over applied voltage *U*. Squares with error bars are experimental data from [4] for nanodomains recorded in thin $LiTaO_3$ film with thickness $h = 55\,nm$; $R_0 = 25\,nm$, $\Delta R \sim 1nm$. Solid line is our fitting calculated at: $R_d = 100\,\mu m$, $P_S \approx 55\,\mu C/cm^2$ [4], $\varepsilon_a = 53$, $\varepsilon_c = 45$, $\varepsilon_e = 81$, $\psi_S \approx 220\,mJ/m^2$, $U_m \approx 6.5\,V$.

Note, that the fitting value of the carrier's emission threshold field $E_m \approx U_m/h = 1180\,kV/cm$ is the same order as perfect $LiTaO_3$ thermodynamic coercive field, but it is significantly higher than the values $E_C \sim (150-200)kV/cm$ measured at 1kHz in the same $LiTaO_3$ plates with thickness 500*nm* [4]. The fitting value of polarization $P_S \approx 55\,\mu C/cm^2$ was measured in the same $LiTaO_3$ samples at 1kHz [4].

The relatively small fitting value of Debye screening radius $R_d = 100\,\mu m$ could be explained by the presence of numerous lithium vacancies reported by authors [4] as well as by the carriers injection from the tip caused by the strong electric field inside the thin film. But the question about the applicability of our "equilibrium" calculations remains open due to small duration of applied voltage pulses $\sim 10^{-4}\,s$. Despite this warning, we obtained rather good quantitative agreement between calculated domain radiuses and recorded ones in thin $LiTaO_3$ films at different applied voltages (see Figs. 8 and 10).

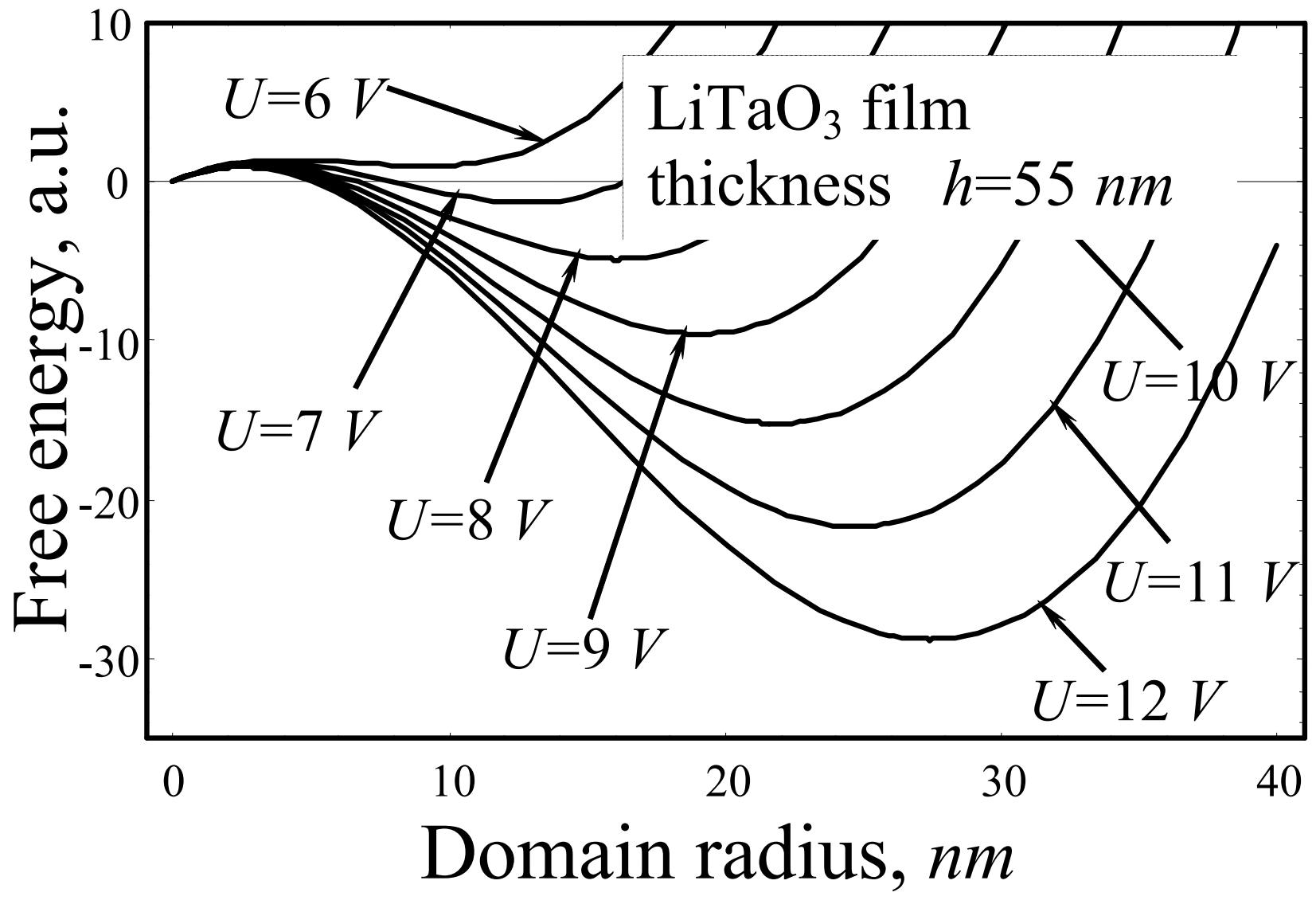

**Figure 9.** Free energy dependence on the domain radius for the different applied voltage values $U$ and the same parameters as in the Fig. 8.

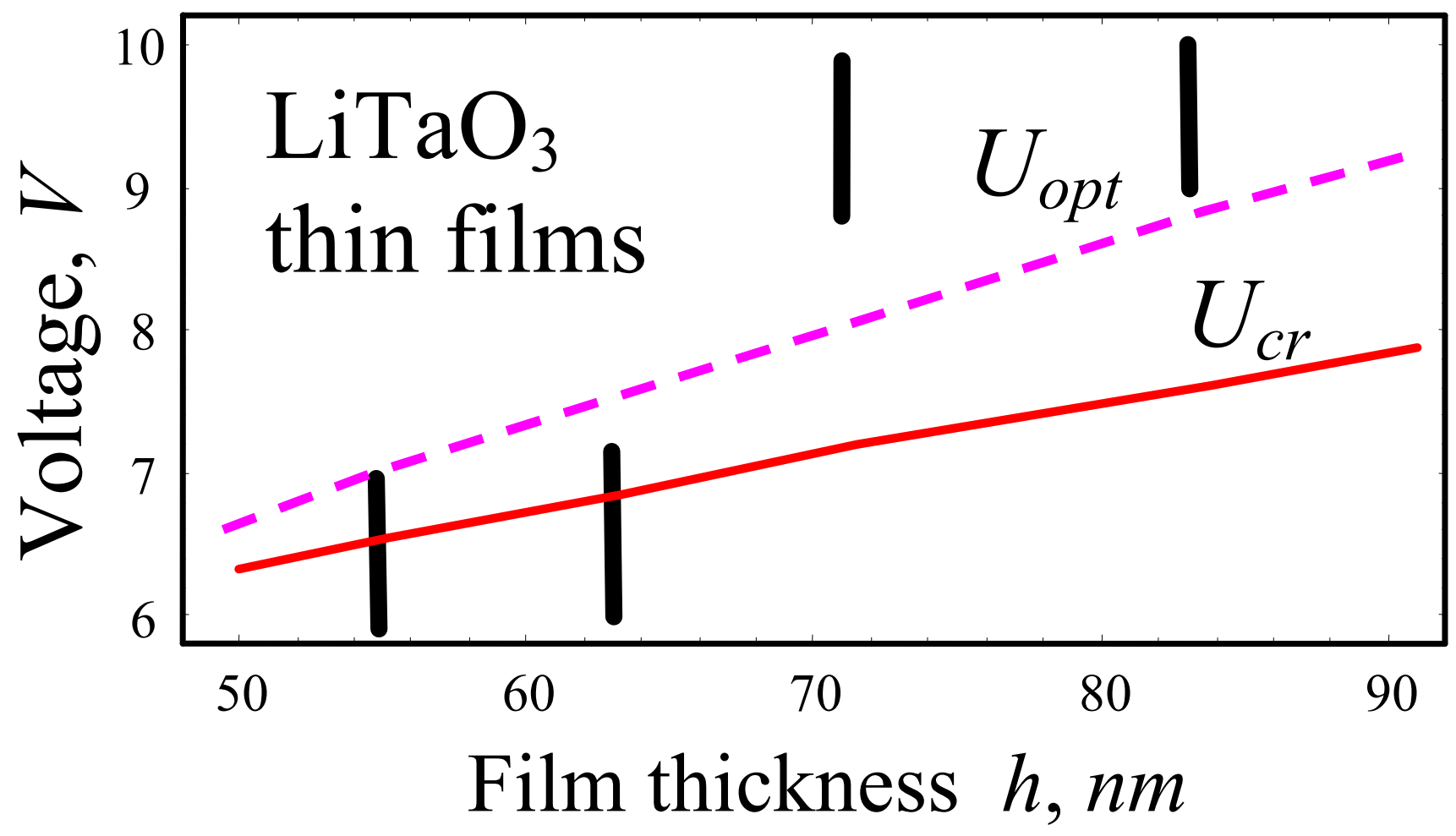

**Figure 10.** The dependences of the critical and optimum voltages over film thickness $h$. Bars are experimental data [4] for the minimum voltage of nanodomains origin. Solid curve for $U_{cr}$ is our fitting calculated from (7) at $E_m \approx 1180\,kV/cm$. Dashed curve for $U_{opt}$ is our modelling for voltage pulses with duration $10^{-4}$ $s$. Other parameters are the same as in the Fig. 8.

## CONCLUSION

- For the first time we obtained the realistic sizes of nanodomains recorded by the electric field of atomic force microscope tip in $BaTiO_3$, $Pb(Zr,Ti)O_3$ and $LiTaO_3$ thin films, in contrast to the over-estimated ones calculated in the previous papers [11], [12], [16].

- First of all we modified the existing models for equilibrium domain formation allowing for Debye screening effects, depolarization energy of the domain butt, bottom electrode influence, recharging of sluggish surface screening layers, field emission between the tip apex and the sample surface originated at high electric fields. We have shown that these effects essentially influence on the equilibrium domain sizes.
- For the first time we demonstrated, that the domain radius does not decrease continuously with voltage decrease: the domain appears with non-zero radius at definite critical voltage. Such "threshold" domain formation is similar to the well-known first order phase transition. We hope that our results will help one to determine the necessary recording conditions and appropriate ferroelectric medium in order to obtain the stable domains with minimum lateral size in a wide range of applied voltages.

**ACKNOWLEDGMENTS**

Authors are grateful to Profs. N.V. Morozovsky and S.L. Bravina for useful discussions of our model.

**REFERENCES.**

## APPENDIX A

For the extrinsic semiconductor with donors concentration $n_d$, the free charges bulk density $\rho_f$ is determined via electric field potential $\varphi(\mathbf{r})$, as follows:

$$\rho_f(\mathbf{r}) = e\left(N n_d \exp\left(-\frac{Ne\varphi(\mathbf{r})}{k_B T}\right) - n_0 \exp\left(\frac{e\varphi(\mathbf{r})}{k_B T}\right)\right) \tag{A.1}$$

Hereinafter we consider the case $\left|Ze\varphi(\mathbf{r})/k_B T\right| << 1$. Taking into account the electro neutrality condition $Z n_d = n_0$, one can find that

$$\rho_f(\mathbf{r}) \approx -\frac{\varepsilon_i\,\varphi(\mathbf{r})}{4\pi R_d^2}, \quad R_d^2 = \frac{\varepsilon_i\, k_B T}{4\pi e^2\left(Z^2 n_d + n_0\right)}. \tag{A.2}$$

The external electric field potential $\varphi_q(\mathbf{r})$ created by the point charge $q$ localized in air in the point $r_0 = (0,0,-z_0)$, inside the film $0 \le z \le h$ filled by isotropic semiconductor with dielectric permittivity $\varepsilon_i$ could be found from the boundary problem:

$$\begin{gathered}
\Delta\varphi_0(\mathbf{r}) = -4\pi\frac{q}{\varepsilon_e}\delta(x,y,z+z_0), \qquad z \le 0, \\
\Delta\varphi_q(\mathbf{r}) - \frac{\varphi_q(\mathbf{r})}{R_d^2} = 0, \qquad 0 \le z \le h, \\
\varphi_0(z=0) = \varphi_q(z=0), \quad \left.\left(\varepsilon_e\frac{\partial\varphi_0}{\partial z} - \varepsilon_i\frac{\partial\varphi_q}{\partial z}\right)\right|_{z=0} = 0, \quad \varphi_q(z=h) = 0
\end{gathered} \tag{A.3}$$

The solution of (A.3) can be found with the help of Hankel integral transformation. Finally we obtained that:

$$\varphi_0(\mathbf{r}) = \frac{q}{\varepsilon_e}\int_0^\infty dk J_0\left(k\sqrt{x^2+y^2}\right)\left(\exp(-k\cdot|z+z_0|) + \exp(k\cdot(z-z_0))\frac{\Delta_-}{\Delta_+}\right) \tag{A.4}$$

$$\varphi_q(\mathbf{r}) = q\int_0^\infty dk J_0\left(k\sqrt{x^2+y^2}\right)\exp(-k z_0)\frac{2k\left(\exp\left(-z\tilde{k}\right) - \exp\left(-(2h-z)\tilde{k}\right)\right)}{\Delta_+} \tag{A.5}$$

$$\Delta_\pm = k\left(1-\exp\left(-2h\tilde{k}\right)\right)\varepsilon_e \pm \tilde{k}\left(1+\exp\left(-2h\tilde{k}\right)\right)\varepsilon_i, \quad \tilde{k} = \sqrt{k^2 + R_d^{-2}}$$

Hereinafter $J_0$ is Bessel function of zero order, $z_0 = R_0 + \Delta R$.

The potential $\varphi_D(\mathbf{r})$ of depolarization field created by the charges on the surface of sample has to satisfy the following system of equations:

$$\begin{gathered}
\Delta\varphi_D(\mathbf{r}) = 0, \qquad z < 0; \\
\Delta\varphi_D(\mathbf{r}) - \frac{\varphi_D(\mathbf{r})}{R_d^2} = 0, \qquad 0 \le z \le h; \\
\varepsilon_e\left.\frac{\partial\varphi_D}{\partial z}\right|_{z=0-0} - \varepsilon_i\left.\frac{\partial\varphi_D}{\partial z}\right|_{z=0+0} = \begin{cases} 4\pi(\sigma_S - P_S), & \sqrt{x^2+y^2} < d \\ 0, & \sqrt{x^2+y^2} > d \end{cases} \\
\varphi_D(z=0+0) = \varphi_D(z=0-0), \qquad \varphi_D(z=h) = 0.
\end{gathered} \tag{A.6}$$

Using the same method as before one can obtain $\varphi_D(\mathbf{r})$ in the form:

$$\varphi_D(\mathbf{r}) = 2\pi(\sigma_S - P_S)\int_0^\infty dk J_0\left(k\sqrt{x^2+y^2}\right)\frac{2J_1(kd)d}{\Delta_+}\begin{cases} \exp(kz)\left(1-\exp\left(-2h\tilde{k}\right)\right), & z \le 0; \\ \exp\left(-z\tilde{k}\right) - \exp\left(-(2h-z)\tilde{k}\right), & 0 < z \le h, \end{cases} \tag{A.7}$$

Hereinafter $J_1$ is Bessel function of first order.

When using the obtained potentials (A.4), (A.5) and (A.7) we can calculate the free energy of the electrostatic field. For the considered case of rigid ferroelectrics it has the following form [24]:

$$\Phi_{el}[\varphi]=\int_V dv\frac{\mathbf{D}\cdot\mathbf{E}-4\pi\mathbf{P}_S\cdot\mathbf{E}}{8\pi}\equiv\int_{z<0}dv\frac{\varepsilon_e\mathbf{E}^2}{8\pi}+\int_{0<z\le h}dv\frac{\varepsilon_i\mathbf{E}^2}{8\pi}\equiv\int_{z<0}dv\frac{\varepsilon_e(\nabla\varphi)^2}{8\pi}+\int_{0<z\le h}dv\frac{\varepsilon_i(\nabla\varphi)^2}{8\pi}\qquad(A.8)$$

Since the self energy of external source does not depend on the domain parameters we found the excess of electrostatic energy $\Delta\Phi_{el}=\Phi_{el}[\varphi_q+\varphi_D]-\Phi_{el}[\varphi_q]$ caused by polarization reversal inside the domain:

$$\Delta\Phi_{el}=\int_V dv\frac{\varepsilon(z)}{8\pi}\left((\nabla\varphi_D)^2+2\nabla\varphi_D\nabla\varphi_q\right)\equiv\Phi_D+\Phi_q\qquad(A.9)$$

Here the first term is the self-energy of depolarization field; the last one is the energy of interaction between the external field and the polarization of the domain. These two summands of the electrostatic energy can be written as follows:

$$\begin{gathered}\Phi_D=(2\pi(\sigma_S-P_S)d)^2\int_0^\infty dk\left(\frac{J_1(kd)}{\Delta_+}\right)^2\times\\ \times\left((1-\exp(-2h\tilde{k}))^2\varepsilon_e+\frac{k}{\tilde{k}}\left(\left(\frac{\tilde{k}^2}{k^2}+1\right)\frac{1-\exp(-4h\tilde{k})}{2}+\left(\frac{\tilde{k}^2}{k^2}-1\right)2h\tilde{k}\exp(-2h\tilde{k})\right)\varepsilon_i\right)\end{gathered}\qquad(A.10)$$

$$\begin{gathered}\Phi_q=4\pi(\sigma_S-P_S)qd\int_0^\infty dk\frac{kJ_1(kd)\exp(-kz_0)}{\Delta_+{}^2}\times\\ \times\left((1-\exp(-2h\tilde{k}))^2\varepsilon_e+\frac{k}{\tilde{k}}\left(\left(\frac{\tilde{k}^2}{k^2}+1\right)\frac{1-\exp(-4h\tilde{k})}{2}+\left(\frac{\tilde{k}^2}{k^2}-1\right)2h\tilde{k}\exp(-2h\tilde{k})\right)\varepsilon_i\right)\end{gathered}\qquad(A.11)$$

Integrals in (A.10) and (A.11) can be integrated in several extreme cases, namely when one of the characteristic lengths which are involved in (A.4), (A.5), (A.7), (A.10), (A.11), is much smaller than others. In this way we obtained the following:

$$\Phi_D=\begin{cases}\dfrac{\pi^2(\sigma_S-P_S)^2}{\varepsilon_i}d^2R_d & at\quad R_d<<\{d,h\}\\ \dfrac{\pi^2(\sigma_S-P_S)^2}{\varepsilon_i}2d^2h & at\quad h<<\{d,R_d\}\\ \dfrac{\pi^2(\sigma_S-P_S)^2}{\varepsilon_i+\varepsilon_e}\dfrac{16}{3\pi}d^3 & at\quad d<<\{R_d,h\}\end{cases}\qquad(A.12)$$

$$\Phi_q = \frac{4\pi(\sigma_S - P_S)q}{\varepsilon_i}\frac{\sqrt{z_0^2+d^2}-z_0}{\sqrt{z_0^2+d^2}}\begin{cases}\dfrac{R_d}{2} & at \quad R_d << \{d,h\}\\ h & at \quad h << \{d,R_d\}\\ \dfrac{\varepsilon_i}{\varepsilon_i+\varepsilon_e}\sqrt{z_0^2+d^2} & at \quad d << \{R_d,h\}\end{cases} \tag{A.13}$$

With help of these expressions we obtained the simplest Pade approximations of 1/1 order that gives all extreme cases of (A.12) and (A.13):

$$\Phi_D \approx \frac{\pi^2(\sigma_S - P_S)^2 d^3 h R_d}{\varepsilon_i dh + \varepsilon_i \dfrac{dR_d}{2} + (\varepsilon_i+\varepsilon_e)\dfrac{3\pi}{16}hR_d} \tag{A.14}$$

$$\Phi_q \approx \frac{4\pi(\sigma_S - P_S)qhR_d\left(\sqrt{z_0^2+d^2}-z_0\right)}{(\varepsilon_i+\varepsilon_e)hR_d + \varepsilon_i(R_d+2h)\sqrt{z_0^2+d^2}} \tag{A.15}$$

For the real tip with radius of curvature $R_0$ and typical recording condition $\Delta R << R_0$, one has to consider not only the effective charge $q = UR_0$, but also all other images in $\varphi_U(\mathbf{r})$ and $\varphi_D(\mathbf{r})$ (see e.g. (A.7), (B.8-9), (C.3) in [13]).

For the case $\Delta R << R_0$ we derived that $q \to UR_0\varepsilon_e \dfrac{\varepsilon_i+\varepsilon_e}{\varepsilon_i-\varepsilon_e}\ln\left(\dfrac{\varepsilon_i+\varepsilon_e}{2\varepsilon_e}\right)$ and $z_0 \to R_0$ in (A.14). For the case $\Delta R << R_0$ and $d \gtrsim R_0$ we derived that $d^3$ in (A.15) should be substituted by: $d^3 \to d^3\left(1-\dfrac{3\pi}{\varepsilon_i+\varepsilon_e}\left(\dfrac{R_0}{d}\right)^2\left(\sqrt{1+\left(\dfrac{d}{R_0}\right)^2}-1+\ln\left(\dfrac{\varepsilon_i+\varepsilon_e}{2\varepsilon_e}\sqrt{1+\left(\dfrac{d}{R_0}\right)^2}-\dfrac{\varepsilon_i-\varepsilon_e}{2\varepsilon_e}\right)\right)\right)$. For the case $\Delta R << R_0$ and $d << R_0$ we derived that $d^3 \to d^2\Delta R\varepsilon_i/\varepsilon_e$ in expression (A.15).

Using the electrostatic energy expressions (A.14), (A.15) and the surface energy of the domain wall one can write the domain free energy in the following form:

$$\Phi(d) \approx \frac{\varepsilon_i+\varepsilon_e}{\varepsilon_i-\varepsilon_e}\ln\left(\frac{\varepsilon_i+\varepsilon_e}{2\varepsilon_e}\right)\cdot\frac{4\pi\varepsilon_e(\sigma_S-P_S)\cdot UR_0\cdot hR_d\left(\sqrt{z_0^2+d^2}-z_0\right)}{(\varepsilon_i+\varepsilon_e)hR_d+\varepsilon_i(R_d+2h)\sqrt{z_0^2+d^2}} + \\ + \frac{\pi^2(\sigma_S-P_S)^2 d^3 hR_d}{\varepsilon_i dh + \varepsilon_i\dfrac{dR_d}{2}+(\varepsilon_i+\varepsilon_e)\dfrac{3\pi}{16}hR_d} + 2\pi\psi_S d\,h \tag{A.16}$$

Then the equilibrium domain radius $d_{eq}$ can be found from the conditions $\partial\Phi(d)/\partial d\big|(d \to d_{eq}) = 0$, $\partial^2\Phi(d)/\partial d^2\,(d \to d_{eq}) > 0$. In the critical point $\Phi(d_{eq} = d_{cr}) = \Phi(d = 0)$ the following condition should be satisfied:

$$U_{cr}(h) = \sqrt{\frac{32\psi_S}{3}\frac{\varepsilon_i R_0 + (\varepsilon_i + \varepsilon_e)h}{\sqrt{(\varepsilon_i + \varepsilon_e)h}}\left(\varepsilon_e \frac{\varepsilon_i + \varepsilon_e}{\varepsilon_i - \varepsilon_e}\ln\left(\frac{\varepsilon_i + \varepsilon_e}{2\varepsilon_e}\right)\right)^{-1}}. \tag{A.17}$$

Here $U_{cr}$ is the critical value of the applied voltage. Only when the applied voltage $U$ exceeds $U_{cr}$ domains are stable. Minimal radius of stable domains is the following:

$$d_{cr} = \sqrt{\frac{3\psi_S}{8(\sigma_S - P_S)^2}(\varepsilon_i + \varepsilon_e)h} << h. \tag{A.18}$$